\documentclass[aps,twocolumn,showpacs,preprintnumbers,floatfix]{revtex4}
\usepackage{graphicx}% Include figure files
\usepackage{dcolumn}% Align table columns on decimal point
\usepackage{bm}% bold math

\begin{document}

\title{Scalar Mesons $a_0(1450)$ and $\sigma(600)$ from Lattice QCD}
\author{N. Mathur$^{a}$, A. Alexandru$^{b}$, Y. Chen$^{c}$, S.J.
Dong$^{b}$, T. Draper$^{b}$, \mbox{I. Horv\'{a}th$^{b}$}, F.X.
Lee$^{d}$, K.F. Liu$^{b}$, S. Tamhankar$^{e}$, and J.B.
Zhang$^{f}$}
\affiliation{$^{a}$Jefferson Lab, 12000 Jefferson Avenue, Newport News, VA 23606 \\
{$^{b}$Dept.\ of Physics and Astronomy, University of Kentucky, Lexington, KY 40506 \\
$^{c}$Institute of High Energy Physics, Beijing 100039, China \\
$^{d}$Dept.\ of Physics, George Washington University, Washington, DC 20052 \\
$^{e}$Dept. of Physics, Hamline University, St. Paul, MN 55104 \\
$^{f}$Dept.\ of Physics, Zhejiang University, Hangzhou, Zhejiang 310027, China}}
\begin{abstract}
We study the $a_0$ and $\sigma$ mesons with the overlap fermion in the chiral
regime with the pion mass as low as $182\,{\rm MeV}$ in the quenched
approximation.  After the $\eta' \pi$ ghost states are separated, we find the
$a_0$ mass with $q\bar{q}$ interpolation field to be almost independent of the
quark mass in the region below the strange quark mass.  The chirally
extrapolated results are consistent with $a_0(1450)$ being the $u\bar{d}$ meson
and $K_0^*(1430)$ being the $u\bar{s}$ meson with calculated masses at $1.42\pm
0.13$ GeV and $1.41\pm 0.12$ GeV respectively.  We also calculate the scalar
mesonium with a tetraquark interpolation field.  In addition to the two pion
scattering states, we find a state at $\sim 550$ MeV.  Through the study of
volume dependence, we confirm that this state is a one-particle state, in
contrast to the two-pion scattering states.  This suggests that the observed
state is a tetraquark mesonium which is quite possibly the $\sigma(600)$ meson.
\end{abstract}
\pacs{12.38.Gc, 14.20.Gk, 11.15.Ha}
\maketitle
\vfill

\section{Introduction}

\begin{figure}
  \centerline{%
    \includegraphics[width=8cm]{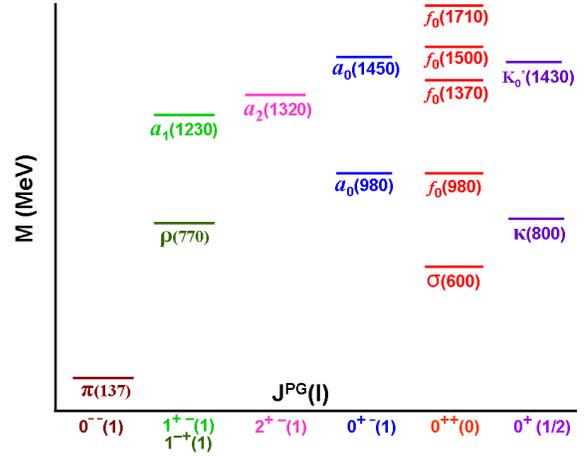}
 }
% \vspace*{-0.1in}
  \caption{\label{scalar} Spectrum of scalar mesons together with
 $\pi$, $\rho, a_1$ and $a_2$ mesons.}
%  \vspace*{-0.2in}
\end{figure}

Unlike pseudoscalar, vector, and tensor mesons, the scalar mesons are not well
known in terms of their $SU(3)$ classification, the particle content of their
composition, or their spectroscopy.  Part of the problem is that there are too
many experimental candidates for the $q\bar{q}$ nonet.  Fig.~\ref{scalar} shows
the current experimentally known scalar mesons whose number more than doubles
that of a nonet.  One viable solution is that low-lying scalars, such as the
$\sigma(600)$, $a_0(980)$ and $f_0(980)$ are tetraquark mesoniums whose
classification and spectroscopy have been studied in the MIT bag
model~\cite{jaf77} and the potential model~\cite{lw81}.  Another suggestion is
that $a_0(980)$ and $f_0(980)$ are $K\bar{K}$ molecular states~\cite{wi82}.
Other candidates for tetraquark mesoniums include vector mesons pairs
produced in $\gamma\gamma$ reactions~\cite{ll82} and hadronic
productions~\cite{ll83} and the recently discovered charmed narrow
resonances~\cite{swa06}.

Under the supposition that $a_0(980)$ and $f_0(980)$ are tetraquark mesoniums
on account of the fact that they are favored by spectroscopy
studies~\cite{jaf77,lw81}, small two-photon decay widths~\cite{bar85}, and the
pattern of $\phi$ and $J/\Psi$ decays~\cite{aks06}, the question remains: where
is the isovector scalar $q\bar{q}$ state?  From Fig.~\ref{scalar}, we see that
one candidate is $a_0(1450)$.  However, in the conventional wisdom of the quark
model, its mass is too high.  Not only is it higher than $a_2(1320)$ and
$a_1(1230)$, in contrast to the spin-orbit splitting pattern in charmonium; it
is even slightly higher than $K_0^{*}(1430)$ which contains a strange quark and
is believed to be the $s\bar{u}$ or $s\bar{d}$ meson in practically all the
models~\cite{pdt04}.  According to the quark counting rule, mesons and baryons
made up of strange quarks are expected to lie higher than their counterparts
with $u/d$ quarks.  Notwithstanding the success of the quark potential model in
describing charm and bottom hadrons, its applicability to light hadrons with
$SU(6)$ symmetry has been questioned, since chiral symmetry plays an essential
role~\cite{ldd99,mcd05} in light hadron dynamics.  Might it be that the scalar
$q\bar{q}$ meson is yet another challenge to the $SU(6)$ quark model's
delineation of light hadrons?

Lattice QCD is perhaps the most desirable tool to adjudicate the theoretical
controversy surrounding the issue and to reveal the nature of the scalar
mesons.  In fact, there have been several calculations to study the $a_0$ meson
with the $\bar{\psi}\psi$ interpolation field in the quenched
approximation~\cite{lw00,bde02,po03} and with dynamical
fermions~\cite{bbo01,pdi04,kun04,gim06,mm06}.  In calculations with relatively
small quark masses, it is found that the $a_0$ mass does not change much below
the strange quark mass.  Consequently, the chiral limit result is consistent
with $a_0(1450)$.  Furthermore, it is emphasized that the would-be $\eta' \pi$
ghost states give negative contributions to the $a_0$ correlator in both the
quenched case~\cite{bde02} and the partially quenched case~\cite{pdi04} when
the quark mass is lower than the strange.  Thus it is essential to take out
these quenched or partially quenched artifacts before one can confidently
obtain the $a_0$ mass.  In the case of $\sigma$, a calculation of the
tetraquark mesonium with the pseudoscalar-pseudoscalar 4-quark interpolation
field~\cite{aj00} has been performed.  From the deviation of the lowest state
mass from that of the expected two-pion scattering state, it was
suggested~\cite{aj00} that a bound mesonium state is seen.

In the present work, we shall use the overlap fermion~\cite{neu98} to calculate
$a_0$ and the scalar tetraquark mesonium.  The overlap fermion has the benefit
of having exact chiral symmetry at finite lattice spacing.  Since chiral
symmetry plays a pivotal role in these mesons in chiral effective theories (in
particular, it is concluded in the recent dispersion analysis of $\pi\pi$
scattering that the occurrence of $\sigma$ is on the basis of ``current
algebra, spontaneous symmetry breakdown, and unitarity''~\cite{cgl01}), we
believe it is desirable to adopt a fermion action which explicitly exhibits the
spontaneously broken chiral symmetry at finite lattice spacing.

For the study of $a_0$, it is well known that there are $\eta' \pi$ ghost
states in the quenched~\cite{bde02,po03} and partially quenched~\cite{pdi04}
cases.  These ghost states become the physical $\eta \pi$ and $\eta' \pi$
states in the full dynamical calculation without partial quenching.  Such
physical two meson states in full QCD have been seen when the quark masses are
light enough~\cite{bbo01}.  Thus, in order to obtain $a_0(1450)$ and possibly
$a_0(980)$ in the quenched approximation, one needs an algorithm which can fit
multiple states including the $\eta' \pi$ ghost states which lie lower in mass
than the $I=1$ scalar $q\bar{q}$ below the strange quark mass region.  We have
developed a sequential empirical Bayes method (SEBM) for constrained-curve
fitting~\cite{ddh03b} to fit multiple states.  This is based on the
constrained-curve fitting of lattice data with Bayesian
priors~\cite{lep02,mor02}.  In the sequential empirical Bayes
method~\cite{ddh03b}, one extracts the priors for the mass and spectral weight
from a subset of data by fitting the two-point correlation function starting
from the large time separation.  First, one fits the ground state in a time
window and then uses its fitted mass and spectral weight as priors to fit the
first excited state in an extended time window.  The process is repeated until
time slices are exhausted.  One then does a constrained-curve fit to the rest
of the data set with the extracted priors.  This method has been employed to
fit the Roper resonance, the radially excited nucleon, and $S_{11}(1535)$ on
top of the $\eta' N$ ghost states~\cite{mcd05}.  It has also been used in
extracting radially-excited states of 1P charmonium~\cite{cll07}.

It turns out that, for the range of quark masses that we fitted, the 3-volume
dependence (from a comparison of $16^3 \times 28$ and $12^3 \times 28$
lattices) of the spectral weights of the respective ghost $\eta' N$
two-particle scattering state and the bound one-particle baryon state come out
in agreement with expectation, as derived in Ref.~\cite{mcd05,mla04}.  We
regard this as a highly non-trivial test for the fitting method.  Similarly, in
the study of the pentaquark state $\Theta^+(1540)$, we fitted the state in
addition to the $\eta' KN$ ghost state and found from the volume dependence
that it is in agreement with the $KN$ scattering state for a large range of
quark masses~\cite{mla04}.  Through these studies, we are more confident that
the fitting method is capable of fitting multiple states including the ghost
states.  It is of course limited by how good the data are and how many states
one can fit, given the number of time slices of the lattice.  We shall use this
algorithm in the present study of $a_0(1450)$ and $\sigma(600)$.  The smallest
pion we have is 182(8) MeV, which is substantially lower than most of the
previous calculations of $a_0$; this allows us to study the behavior of $a_0$
with the pseudoscalar meson mass ranging from 1.3 GeV down to 182 MeV.  This is
important in revealing that the $a_0$ mass is very insensitive to quark mass in
the range from strange quark down to the physical $u/d$ mass.  This turns out
to have significant phenomenological implications on the pattern of scalar
mesons~\cite{ccl06a}.  For the study of the $\pi\pi$ mesonium with the
tetraquark interpolation field on our lattice, it is crucial for the pion mass
to be lower than $\sim 300$ MeV in order to be able to disentangle the $\pi\pi$
scattering states from the one-particle mesonium, in order to reveal the nature
of the fitted states from the tetraquark correlator.  Thus, for both the case
of $a_0(1450$ and $\sigma(600)$, it is essential to study them in the chiral
regime where $m_{\pi}$ is smaller than $300$ MeV.

\section{$a_0(1450)$ and $K_0^*(1430)$ Mesons}

Our calculation is based on data from $16^3 \times 28$ and $12^3 \times 28$
lattices with 300 quenched Iwasaki gauge configurations ($\beta = 2.264$) and
overlap fermions with a lattice spacing $a = 0.200(3)\,{\rm fm}$ determined
from $f_{\pi}(m_{\pi})$.  This makes our lattice sizes at $3.2\,{\rm fm}$ and
$2.4\,{\rm fm}$, respectively.  We will discuss the scale determination from
the Sommer scale $r_0$ later to assess the systematic error in scale setting.
A subset of these quark propagators was used to study the quenched chiral logs
in pion and nucleon masses~\cite{ddh03a}, the Roper and $S_{11}$~\cite{mcd05},
and the pentaquarks~\cite{mla04}.

There has been a concern that lattice spacing of $0.2$ fm may be too coarse for
the overlap fermion and speculation that the range of the overlap Dirac
operator may be as long as 4 lattice units~\cite{gss05}.  However, direct
calculation~\cite{dmz06} at lattice spacings of 0.2, 0.17, and 0.13 fm with
Iwasaki gauge action reveals that the range of the operator is comfortably
small in each of these cases (one lattice unit in Euclidean distance and 2
units in ``taxi driver'' distance, the latter being defined as 
$ r_{\rm TD} \equiv || x-y||_1 = \sum_{\mu=1,4} |x_\mu - y_\mu|$) and it approaches 
zero toward the continuum limit.  Thus, we don't think there is an issue regarding 
locality of the overlap operator at 0.2 fm that we base our results on.

\begin{figure}[ht]
% \vspace*{-0.05in}
  \centerline{%
  \includegraphics[width=1.0\hsize]{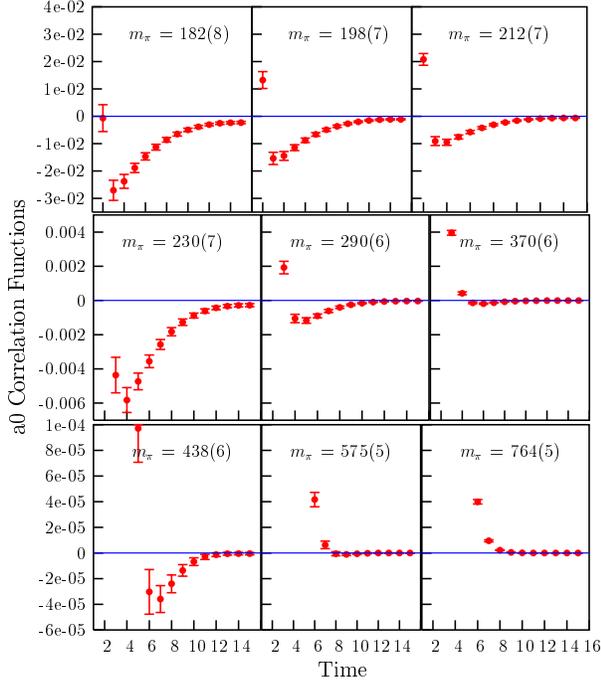}
  }
% \vspace*{-.1in}
  \caption{$a_0$ correlators from the $\bar{\psi}\psi$ interpolation field for
           several quark masses with corresponding pion masses in MeV.}
% \vspace*{-0.1in}
\label{a0-corr}
\end{figure}

We first report results on $a_0$ for which we use $\bar{\psi}\psi$ as the
interpolation field.  Since $a_0$ is an isovector, one only needs to calculate
the correlator with the connected insertion.  Shown in Fig.~\ref{a0-corr} are
$a_0$ correlators as a function of time for 9 quark cases with pion mass from
$m_{\pi} = 182(8)\,{\rm MeV}$ to $m_{\pi} = 764(5)\,{\rm MeV}$.  It is seen
that for pion mass lower than $\sim 600\,{\rm MeV}$, the $a_0$ correlator
starts to develop a negative tail, and it is progressively more negative at
earlier time slices for smaller quark masses.  This is a clear indication that
at least one of the ghost $\eta' \pi$ states, being lightest in mass, are
dominating the correlator over the physical $a_0$ at larger time slices.  This
has been reported in the literature for the quenched~\cite{bde02,po03} and
partially quenched~\cite{pdi04} calculations and the ghost contribution has
been removed with the help of re-summed hairpin diagrams.  The ghost $\eta'\pi$
contribution in the $a_0$ correlator has been studied in the chiral
perturbation theory~\cite{bde02,pdi04}.  The one-loop hairpin diagram gives the
following contribution to the isovector scalar meson propagator
\begin{equation}
\Delta_h(p)= \frac{1}{V_3T}\sum_k \frac{-4 r_0^2 m_0^2}{((p-k)^2 + m_{\pi}^2)
(k^2 + m_{\pi}^2)^2},
\end{equation}
where $V_3=L_x L_y L_z$ is the three-volume of the lattice, $r_0$ is the
coupling between the scalar interpolation field and the $\eta'$ and
$\pi$~\cite{bde02,pdi04} or the matrix element $\langle 0|\bar{\psi}\psi|\eta'
\pi\rangle$, and $m_0^2= 2N_f\,\chi_t/f_{\pi}^2$ is the hairpin insertion mass
which is related to the topological susceptibility $\chi_t$ in the pure gauge
theory.

The Fourier transform (FT) of the dimensionless $\Delta_h(p)$ for the case of
$\vec{p}=0$ gives the following contribution to the scalar meson correlator on
the Euclidean lattice
\begin{eqnarray}  \label{G_S}
G_S(\vec{p}=0) = FT \{a^2\Delta_h(\vec{p}=0)\} &=& - \frac{r_0^2m_0^2N_T}{2N_S^3} \times \nonumber\\
&& {\hspace{-1.8in}}\sum_{\vec{k}} \frac{(1+E_{\pi}t)}{2E_{\pi}^4} e^{-2E_{\pi}t} + 
(t \longrightarrow N_T -t),
\end{eqnarray}
where $E_{\pi} = \sqrt{\vec{k}^2 + m_{\pi}^2}$ and the $(1+E_{\pi}t)$ factor is
due to the double pole of the would-be $\eta'$ ghost propagator in the loop.
$N_S$ and $N_T$ are the number of lattice points in the spatial and time
direction respectively.
The result for the more general partially quenched case has been derived in
Ref.~\cite{pdi04} and the corresponding expression for the would-be $\eta' - N$
one-loop contribution to the nucleon correlator is derived in the study of
Roper and $S_{11}(1535)$~\cite{mcd05}.

We shall use the expression
\begin{equation}   \label{form_ghost}
W(1 + E_{\pi} t) e^{- E_{\eta' \pi}t},
\end{equation}
where $W$ is referred to as the spectral weight in later discussion, to model
the fit of each ghost state contribution.  Here we allow the energy, $E_{\eta'
\pi}$, of the interacting $\eta'-\pi$ to be fitted to the data, but retain the
double pole character of the prefactor $1 + E_{\pi} t$.  This should be a good
approximation when the $\eta' - \pi$ interaction is weak (N.B.  in the large
$N_c$ consideration, the meson-meson interaction is of the order $1/N_c$) so
that the prefactor due to the double pole in the would-be $\eta'$ propagator
remains largely valid when higher orders are included.  We have used a similar
expression to fit the $\eta' N$ ghost states in the nucleon and $S_{11}(1535)$
correlators~\cite{mcd05} and found that their spectral weights have the correct
3-volume dependence for the range of quark masses we calculated.  Thus, we
shall employ this form to fit the $a_0$ correlator and, as a cross check, will
examine if the $1/E_{\pi}^4$ dependence in Eq.~(\ref{G_S}) is borne out from
the fit.

\begin{figure} [htb]
%\vspace*{-0.08in}
  \centerline{%
  \includegraphics[width=1.0\hsize]{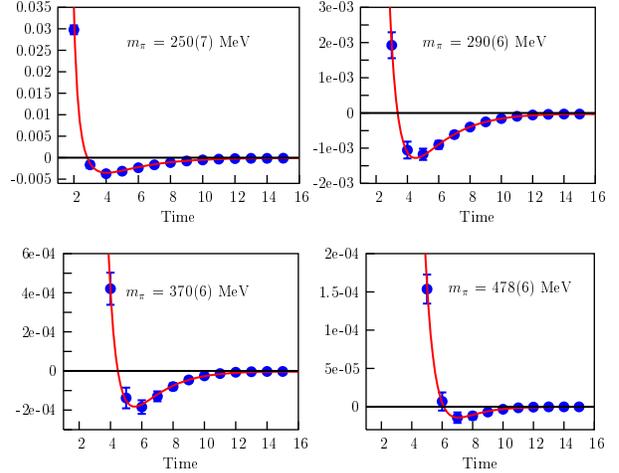}
  }
% \vspace*{-0.16in}
  \caption{Fitted $a_0$ correlators with a low-lying ghost $\eta'\pi$ state for
           several quark masses.   }
% \vspace*{-0.2in}
\label{a0fit}
\end{figure}

We use the above-mentioned SEBM~\cite{ddh03b} to perform the curve-fitting with
the weight $W$ of Eq.~(\ref{form_ghost}) constrained to be negative, to reflect
the ghost nature of the state as shown in Fig.~\ref{a0-corr}, and the total
energy of the would-be $\eta'$ and $\pi$ constrained to be not far from the
energy of the two non-interacting pions, i.e.\ $2\sqrt{p_{n}^2 + m_{\pi}^2}$
with discrete lattice momenta $p_n = \frac{2\pi n}{L}, n = 0, \pm 1, \cdots$.
In the course of studying SEBM~\cite{ddh03b}, we extracted the priors from a
subset of the data and applied them in a constrained fit of the rest of the
data; we found that the results were very compatible, without detectable bias,
with those obtained when the priors were applied to the full data set.  In the
present work, we extracted the priors from 100 gauge configurations and applied
them to a fit of the remaining 200 configurations; we find that the results are
very close to those obtained when the priors were applied to the full 300
configurations, except the latter had smaller errors.  We shall report the
results based on the 300 configurations here.  For $m_{\pi} \ge 250$ MeV on the
$12^3 \times 28$ lattice, we have been able to fit 4 states with the lowest and
the third one being the ghost $\eta' \pi$ states which are close to the
non-interacting pair with each meson at zero momentum, and one unit of lattice
momentum ($p_1$), respectively.  The second state has a positive weight and is
interpreted as the physical $a_0$ with the usual exponential form $e^{-Et}$.
We show in Fig.~\ref{a0fit} the fits to the $a_0$ correlators for several low
quark masses.

\begin{figure} [htb]
%\vspace*{-0.08in}
  \centerline{%
  \includegraphics[width=1.0\hsize]{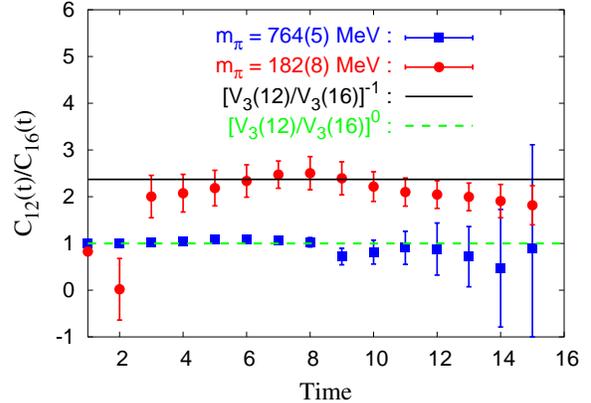}
  }
% \vspace*{-0.16in}
  \caption{Ratio of $a_0$ correlators at $12^3$ and $16^3$ lattices for 
 the cases of $m_{\pi} =
764(5)$ MeV and $m_{\pi}=182(8)$ MeV. While the expected two-particle ghost
$\eta'\pi$ state shows volume dependence at lower quark mass, one particle $a_0$ state
does not show any volume dependence at higher quark mass.}
% \vspace*{-0.2in}
\label{ratio_a0_corr}
\end{figure}

Due to the fact that there are expected to be two to three ghost states below
the $a_0$ on the $16^3 \times 28$ lattice, we have, unfortunately, not been
able to fit all of them to extract the physical $a_0$ with the SEBM fitting in
this case.  There are not enough time slices to fit 5 states.  Otherwise, we
could have compared the spectral weights from the $12^3 \times 28$ and $16^3
\times 28$ lattices and checked the expected volume dependence of the
one-particle $a_0$ and two-particle $\eta'-\pi$ states.  However, we could and
did compare the ratio of the $a_0$ correlators $C_{12}(t)/C_{16}(t)$ between
the $12^3 \times 28$ and the $16^3 \times 28$ lattices.  As was derived in the
course of studying pentaquark states~\cite{mla04}, the spectral weight of a
one-particle state in the point-sink and point-sink correlator is proportional
to unity; whereas, that of a weakly interacting two-particle state is
proportional to $1/V_3$.  We show the ratio of the $a_0$ correlators
$C_{12}(t)/C_{16}(t)$ in Fig.~\ref{ratio_a0_corr} for the cases of $m_{\pi} =
764(5)$ MeV and $m_{\pi}=182(8)$ MeV.  In the case of the heavier pion, the
ratio of the correlators for the whole time range is close to unity.  This
reflects the fact that there are no ghost states, so the lowest state is just
the scalar $q\bar{q}$ meson and the ratio of the spectral weights is
independent of the volume.  On the other hand, the ratio for the
$m_{\pi}=182(8)$ MeV case is close to $[V_3(12)/V_3(16)]^{-1}=2.37$ for $t \ge
3$, indicating that the lowest state is the expected two-particle ghost
$\eta'\pi$ state.

\begin{figure} [htb]
% \vspace*{-0.08in}
  \centerline{%
    \includegraphics[width=0.55\hsize,height=0.55\hsize]{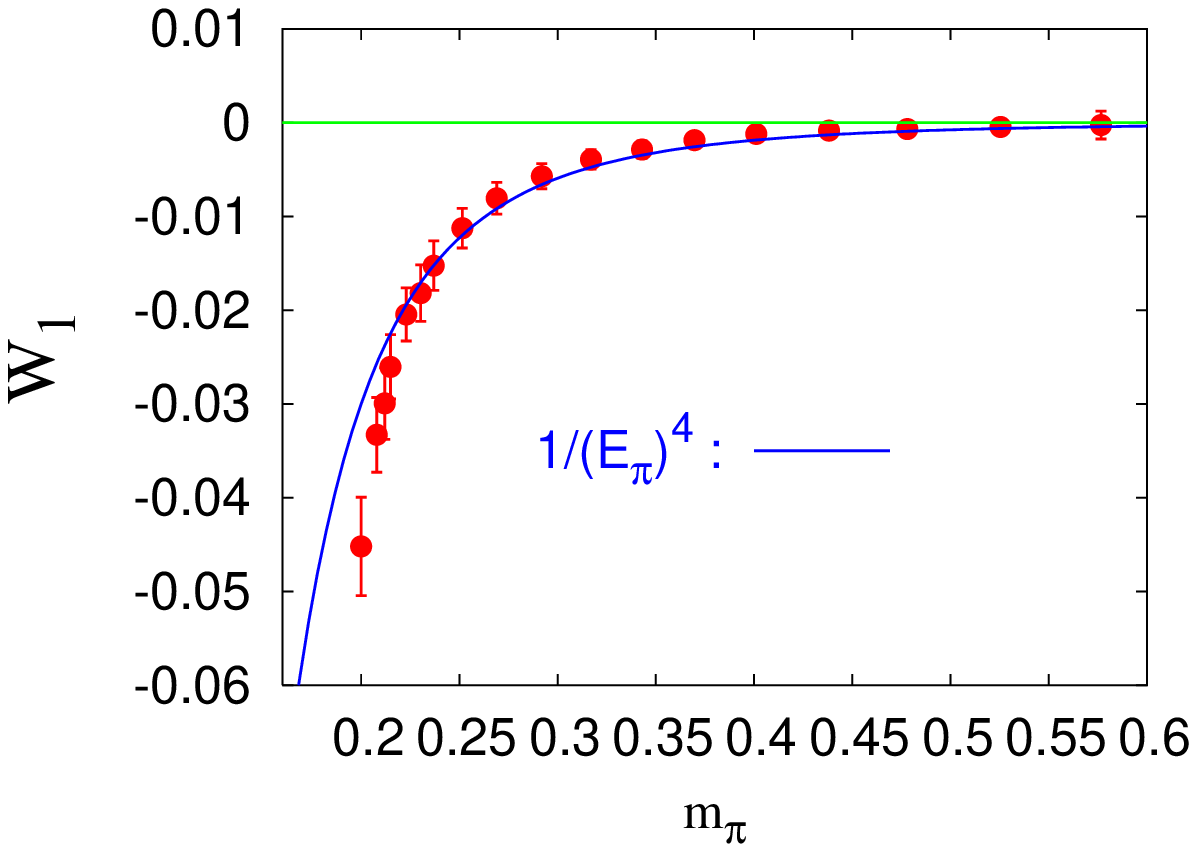}
    \includegraphics[width=0.55\hsize,height=0.55\hsize]{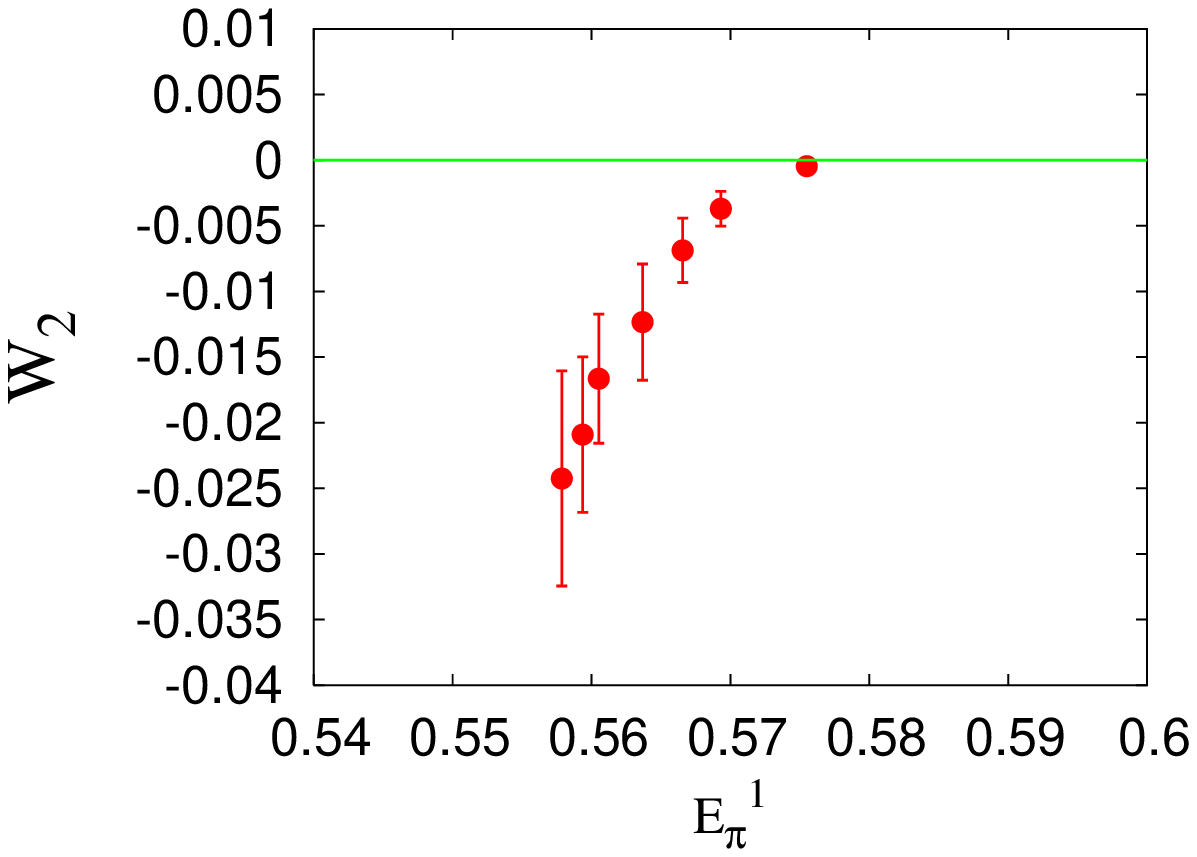}
  }
% \vspace*{-0.16in}
  \caption{Fitted spectral weight $W_1$ for the lowest ghost state (left panel)
           and $W_2$ for the second ghost state in the $a_0$ correlator from
           the $12^3 \times 28$ lattice as a function of the pion mass and
           energy in GeV.   }
% \vspace*{-0.2in}
\label{W_g}
\end{figure}

In addition, we can check the $1/E_{\pi}^4$ dependence in the spectral weight
$W$ as suggested in Eq.~(\ref{G_S}).  We adopted the fitting form in
Eq.~(\ref{form_ghost}) based on the premise that the higher loop diagrams are
not important.  If the spectral weight indeed exhibits the $1/E_{\pi}^4$
dependence, it would lend support for such an assumption.  We note that there
are two ghost $\eta' \pi$ states below the physical $a_0$ on the $12^3 \times
28$ lattice for pion mass below 250 MeV.  In this case, the lowest interacting
would-be $\eta'$ and $\pi$ scattering state is close to the non-interacting
pair with each meson at zero momentum; the second is close to the 
non-interacting pair with each meson having one unit of lattice momentum, i.e.\
$p_1$.  We have been able to fit 4 states in the time window from $t = 12-13$
to $t = 2$.  The third state has a positive weight and is interpreted as the
physical $a_0$ with the usual exponential form $e^{-Et}$.  Plotted in the left
panel of Fig.~\ref{W_g} is the spectral weight $W_1$ that we fitted for the
lowest ghost $\eta'-\pi$ state as a function of the pion mass.  We see that as
the $m_{\pi}$ decreases, it is quite singular.  We fitted with $1/E_{\pi}^4$
from the pion mass from $575$ MeV down to 200 MeV.  It is found that one can
obtain a good fit down to $m_{\pi} \sim 270$ MeV.  Below $\sim 270$ MeV, there
is a deviation.  This is presumably due to the higher order effect.  Similarly,
we plot the spectral weight $W_2$ in the right panel in Fig.~\ref{W_g}. We
see that it is non-zero below $m_{\pi} \sim 250$ MeV and its magnitude
increases as $E_{\pi}$ decreases.  However, it does not increase nearly as fast
as $1/E_{\pi}^4$.  We conjecture that we may not have isolated the second ghost
state in our SEBM fit and it may have included higher ghost contribution.
Since the second ghost state starts to show up below $m_{\pi} = 250$ MeV where
the fitted spectral weight $W_1$ starts to deviate from the expected
$1/m_{\pi}^4$ behavior, we do not quote results on $a_0$ below $m_{\pi} = 250$
MeV.

\begin{figure} [htb]
% \vspace*{-0.08in}
  \centerline{%
  \includegraphics[width=1.0\hsize]{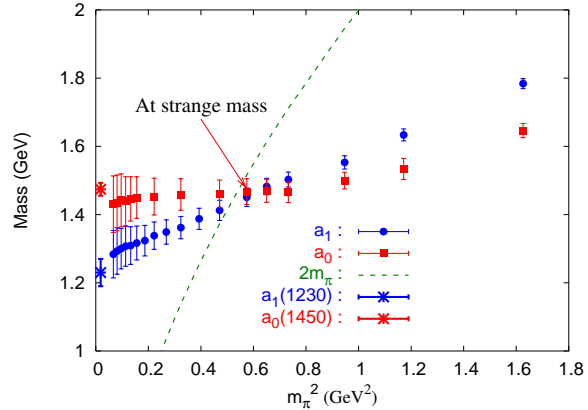}
  }
% \vspace*{-0.16in}
  \caption{$a_0$ and $a_1$ masses are plotted as a function of
           $m_{\pi}^2$.   Also shown is the two pion mass (dashed line) which
           becomes lower than $a_0$ around the strange quark mass region.   }
% \vspace*{-0.2in}
\label{a0_a1}
\end{figure}

Given the caveats of the ghost state fitting, we interpret the second state
from our resultant fit from the $12^3 \times 28$ lattice which has the ordinary
exponential form in time and positive spectral weight to be the physical $a_0$.
We plot its mass as a function of the corresponding $m_{\pi}^2$ in
Fig.~\ref{a0_a1} together with that of $a_1$ for comparison.  The latter does
not have ghost state contamination and is thus easier to calculate.  We see
that above the strange quark mass, $a_1$ lies higher than $a_0$ as expected
from the quark model of heavy quarks.  However, when the quark mass is smaller
than that of the strange, $a_0$ levels off, in contrast to $a_1$ and the other
hadrons that have been calculated on the lattice.  This confirms the trend that
has been observed in earlier works at higher quark
masses~\cite{lw00,bde02,pdi04}.  The chirally extrapolated mass $a_0 = 1.42 \pm
13$ GeV suggests that the meson $a_0(1450)$ is a $q\bar{q}$ state.  By virtue
of the fact that we do not see $a_0(980)$ which lies lower than $a_0(1450)$, we
estimate the spectral weight ratio $\langle 0|\bar{\psi}\psi|a_0(980)\rangle^2/
\langle 0|\bar{\psi}\psi|a_0(1450)\rangle^2$ to be less than 0.015 from the 
relative error of the correlator in the time window where $a_0$ is fitted.  We
also calculated the $K_0^{\*}(1430)$ mass with the strange mass fixed at $ma=
0.26833$ which gives a vector meson mass corresponding to the $\phi$ mass and
the $u/d$ is extrapolated to the chiral limit.  In this case, we also need to
fit and remove the ghost $K\eta'$ states due to the hairpin diagram of the
$u/d$ quark.  There is no ghost state due to the $s$ quark, since according to
Fig.~\ref{a0-corr}, no ghost state is seen for quark mass greater than 1/4 of
that of the strange.  As a result, we obtain the $K_0^{\*}(1430)$ mass at $1.41
\pm 0.12$ GeV and the corresponding scalar $\bar{s}s$ state from the connected
insertion to be $1.46 \pm 0.05$ GeV.  Our findings are quite consistent with
the experimental fact that $K_0^{\*}(1430)$ is basically degenerate with
$a_0(1450)$ despite having one strange quark.  This unusual behavior is not
understood as far as we know and it serves as a challenge to the existing
hadronic models.

It is known that the scale in the quenched approximation is not determined
uniquely.  We note that if the Sommer scale $r_0=0.5$ fm is used for the scale,
the lattice spacing will be 12\% smaller, i.e.\ $a=0.175(3)$ fm.  As a result,
all the masses determined above will be $\sim14\%$ higher in the $r_0 = 0.5$ fm
scale.  The same is true with the following calculation of the $\pi\pi$ state
and the tetraquark mesonium.

Since we do not see $a_0(980)$ in the $a_0$ correlator with the
$\bar{\psi}\psi$ interpolation field, it leaves room for it to be something
other than a $q\bar{q}$ state, e.g.\ a $q^2\bar{q}^2$ state as suggested in
model studies.  However, it is a challenge to verify it on the lattice due to
the complication that there is a threshold $K\bar{K}$ state nearby (within 10
MeV).  Therefore, we shall study the $\sigma(600)$ first which, if present as a
tetraquark mesonium, is several hundred MeV above the $\pi\pi$ threshold and
several hundred MeV below the next $\pi\pi$ scattering state with momentum
close to $p_1$.  This is so, provided that the pion mass is lower than $\sim
250$ MeV, and it may present the best hope of detecting such a state without
the worry of entanglement with the collateral two-meson scattering states.
Since the lowest pion mass in our case is $182$ MeV, we are in a position to
examine it.

\section{$\sigma(600)$ Meson}

The $\sigma$ meson was first postulated by M.H. Johnson and E. Teller as a
classical field to explain the saturation properties and binding energies of
nuclei; they estimated a mass $\sim 500$ MeV from the surface
energy~\cite{jt55}.  It has been suggested that the $\sigma$ is partially
responsible for the enhancement of the $\Delta I = 1/2$ decay in $K \rightarrow
\pi\pi$~\cite{pha86}.  Although it has been put back in the particle data table
on account of the $D^+ \longrightarrow \pi^-\pi^+\pi^+ $
experiment~\cite{ait01}, its experimental existence is still not fully settled
due to the complication that its large width is as large as its mass.  Recent
dispersion analysis~\cite{ccl06} using the Roy equation has produced a
resonance pole in $\pi\pi$ scattering with high precision.  The mass is given
as $441^{+16}_{-8}$ MeV with a width of $544^{+18}_{-25}$ MeV.  Lattice QCD, in
principle, is capable of resolving the issue about its existence.

Resonance can be viewed as a mixture of a bound state and the scattering states
in the usual potential model description of scattering and resonance. In a
coupled channel approach, one can couple a bound state in the continuum with
the scattering states via a coupling potential resulting in a bound state
leaking to the continuum with a shift in mass and acquiring a width.  On a
hypercubic lattice with periodic boundary conditions, the available
momenta are $p_n= n \frac{2\pi}{La}, n =0, \pm 1, \pm 2 \cdots$ and therefore
the one- and two-meson spectra for a certain quantum number are discrete.
Imagine one considers a very large box where the two-meson spectrum has
closely-spaced levels.  When one looks at the spectral weights of the
correlator in the channel with a specific quantum number, there will be
envelopes of states with enhanced spectral weights which are the result of the
mixture of a bound state and the nearby scattering states.  They are the finite
box representation of the resonances in the continuum.  The ``width'' of the
structure would reflect how far apart in energy the mixing takes place.
If we start to decrease the size of the box, the spectrum is going to be less
dense and states are further apart from each other.  The number of states under
the envelope diminishes.  When the box is small enough so that the scattering
states are spaced enough apart that none is expected to lie under the envelope,
then only the bound state remains and thus can be identified as such.  By being
``far enough apart'' we mean that we can define the separation of the scattering states
$\Delta$ to be several times greater than the width $\Gamma$ of the structure,
i.e.\ $\Delta \gg \Gamma$.  In this case, the bound state and the scattering
states are not mixed and they will each have a different volume dependence in
their spectral weights.  Comparing the spectral weights of the same unmixed
state in two lattice volumes is an effective way of revealing the one- or
two-particle nature of the state~\cite{mcd05,mla04} when the interpolation
field projects to both the bound and scattering spectra in the correlator with
a definite quantum number.  This is our approach of identifying, through the
volume study of their spectral weights, both a bound tetraquark state (the
$\sigma$) and a two-pion scattering state which are reasonably well separated
and unmixed.

Experimentally, $\sigma$ is a very broad state with a width of $544$ MeV
according to the recent dispersion analysis~\cite{ccl06}.  It is natural to ask
if it is ever possible to delineate it with a lattice calculation in Euclidean
space.  To answer this, we shall reverse the above discussion on the existence
of the bound state.  Suppose one finds a bound state in addition to the
scattering states which are outside the ``width'' of the state at a relatively
small volume.  To gain information about the scattering phase shift and hence
the real width, we now increase the box size.  As the box size is increased,
the energies of the scattering states will be lowered since the value of the
discrete momenta decreases.  When the scattering state above the bound state is
lowered to within the range of the ``width'', it mixes with the bound state
(this actually defines the range of the width) and the two states avoid the
level crossing.  From the energy of the scattering state one can deduce the
scattering phase shift using the L\"{u}scher
formula~\cite{lus86,lus91,gl92,rg95,pt06}.  This is valid for elastic
scattering irrespective of how broad the state is.  This is studied in detail
in a 2-D lattice model~\cite{gl92} and a spin model~\cite{rg95} which
illustrate how the scattering state mixes with the bound state and gives rise
to the phase shift as the volume is increased.  In a sense, by varying the
lattice volume, hence the momentum, one can use a scattering state to mix with
the bound state and scan the spectrum to obtain the phase shift and therefore
the width of the resonance.  The information of the width can also be obtained
by determining how far apart in energy the scattering and bound state start to
avoid the level crossing.

In the present manuscript, we are only concerned about the existence of
$\sigma$ and will leave the study of its width to the future.  In this vein, we
have chosen the lattice size such that the two lowest $\pi\pi$ scattering
states are expected to be more than half of the experimental width, i.e.\ $272$
MeV, away from the expected $\sigma$ mass at $\sim 600$ MeV.  If this expected
result is borne out, we can use the volume test to discern the particle content
of the states and thereby distinguish the bound $\sigma$ from the two-pion
scattering states.  This is basically our strategy to seek the existence of
$\sigma$.

\begin{figure}[t]
\vspace*{-0.1in}
  \centerline{%
% \hspace*{+0.04\hsize}
  \includegraphics[width=1.0\hsize]{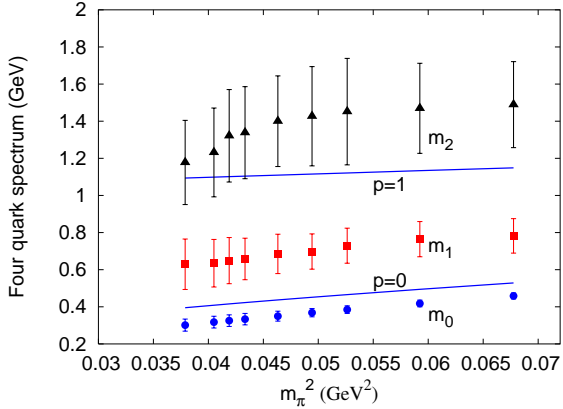}
  }
% \vspace*{-0.1in}
  \caption{The lowest three states from the scalar tetraquark correlator as a
           function of $m_{\pi}^2$ for $m_{\pi}$ from 182 MeV to 250 MeV on the
           $12^3 \times 28$ lattice.  The solid lines indicate the energies of
           the two lowest non-interacting pions in S-wave with lattice momenta
           $p_0=0$ and $p_1 = 2\pi/La$.   }
%  \vspace*{-0.3in}
\label{sigma_pipi}
\end{figure}

To confirm the existence of $\sigma(600)$ beyond doubt, one needs to identify
both the tetraquark mesonium and the collateral $\pi\pi$ scattering states.
Secondly, one needs to work on a lattice where the scattering states and the
bound state are well separated (e.g.\ further apart than half of the width of
the ``would-be'' resonance) to avoid admixture; this is in order to discern the
nature of these states separately to make sure that $\sigma$ is indeed a
one-particle state, not a two-particle scattering state.  To this end, we used
the adaptive Bayes curve-fitting method~\cite{ddh03b} as described above to fit
the ground state and the excited states of the tetraquark correlator with a
local interpolation operator $\bar{\psi}\gamma_5\psi \bar{\psi}\gamma_5\psi$
for both the source and the sink.  We note that it does not matter what
interpolation field one uses for the calculation, as long as it has overlap
with the states of the corresponding quantum number.  Being local,
the pseudoscalar-pseudoscalar operator that we adopt will have vector-vector,
scalar-scalar, $\cdots$ components after Fierz transform.  This aspect has been
discussed in the calculation of the exotic $I=2$ $\rho\rho$
states~\cite{lll90}.  It is also shown in the pentaquark study that different
interpolation fields are related through Fierz transform~\cite{mla04} and the
masses from these interpolation fields were verified to be the same in a
lattice calculation~\cite{lm06}.  Unlike in the $a_0$ correlator, there are no
$\eta' \pi$ ghost states to worry about in this tetraquark channel.  As in
Ref.~\cite{aj00}, we consider only the connected insertion, not the single and
double annihilation insertions.  They are likely to preferentially project to
the higher $q\bar{q}$ and glueball states~\cite{aj00}.  To verify this, we use
two point sources at $t=0$ and $t = 8$ and a zero-momentum wall source in the
Coulomb gauge at $t = 14$ to calculate the disconnected-insertion correlator at
time separations $t - t_0 = 0, 6$, and 14 and found that they are an order of
magnitude smaller than the corresponding connected-insertion correlator.  This
shows that the annihilation diagrams are not likely to change the results of
the connected insertion qualitatively.  We present our results on the $12^3
\times 28$ lattice in Fig.~\ref{sigma_pipi} as a function of $m_{\pi}^2$ for
the pion mass range from 182 MeV to 250 MeV.  We have fitted three states so
that we can trust the results of the two lower ones.  The lowest one is about
$100(20)$ to $60(10)$ MeV below the $\pi\pi$ threshold.  This is most likely
the interacting state of two pions with energy close to and below that of two
non-interacting pions at rest, since the interaction is attractive in the $I
=0$ channel. It is shown that the energy shift of two interacting particles in
a finite box can be related to the infinite volume scattering length below the 
inelastic threshold in a systematic $1/L$ expansion~\cite{lus86}. In particular,
for two spinless bosons with mass $m$ at rest, one has
\begin{equation}  \label{full_QCD}
\Delta E = E- 2 m = -\frac{4\pi a_0}{m L^3}\Biggl(1 + c_1\frac{a_0}{L} + 
c_2 (\frac{a_0}{L})^2 \Biggr)
\end{equation}
where $c_1= -2.873$ and $c_2=6.3752$ from one-loop calculation~\cite{lus86}. However, it is
pointed out~\cite{bg96} that there are would-be $\eta'$ hairpin diagram contributions at the
one-loop order for $\pi\pi$ scattering which spoils the relation between 
$\Delta E$ and the scattering length $a_0$ in Eq.~(\ref{full_QCD}). It has $L^0$ and
$L^2$ terms in the $I=0$ channel in addition to the leading $1/L^3$ term in 
Eq.~(\ref{full_QCD}). Since our calculation is done in the quenched approximation,
using the full QCD one-loop chiral perturbation formula~\cite{lus86}
to extract the scattering length of $\pi\pi$ scattering from the energy shift in the 
finite box channel is not applicable. We shall, instead, compare our results to that
derived in quenched chiral perturbation theory. 
The quenched one-loop $\pi\pi$ scattering energy shift in the
finite box which includes the hairpin diagrams has been derived~\cite{bg96}.
The energy shift is
\begin{equation} \label{energy-shift}
\Delta E = E  - 2 m_{\pi}= \Delta E^{\rm tree} + \Delta E^{\rm one-loop}
\end{equation}
where the tree-level result for the $I=0$ channel is
\begin{equation}   \label{tree}
\Delta E^{\rm tree} = \frac{-7}{4f_{\pi}^2 L^3},
\end{equation}
with $f_{\pi} = 132$ MeV at the physical pion mass and the one-loop result is
given~\cite{bg96} as
\begin{eqnarray}
\Delta E^{\rm one-loop} &=& m_{\pi}\Biggl[B_0(m_{\pi}L)\,\delta^2 
+ A_0(m_{\pi}L)\,\delta\epsilon \nonumber\\
&&\hspace{0.4in} +\, O\Biggl(\frac{\epsilon^2}{(m_{\pi}L)^3}\Biggr)\Biggr],
\end{eqnarray}
where
\begin{equation}
\delta \equiv \frac{m_0^2/3}{8\pi^2 f_{\pi}^2}, \hspace{2cm}
\epsilon \equiv \frac{m_{\pi}^2}{16\pi^2 f_{\pi}^2}.
\end{equation}

\begin{figure}[t]
%\vspace*{-0.1in}
  \centerline{%
% \hspace*{+0.04\hsize}
  \includegraphics[width=0.6\hsize,height=0.15\vsize]{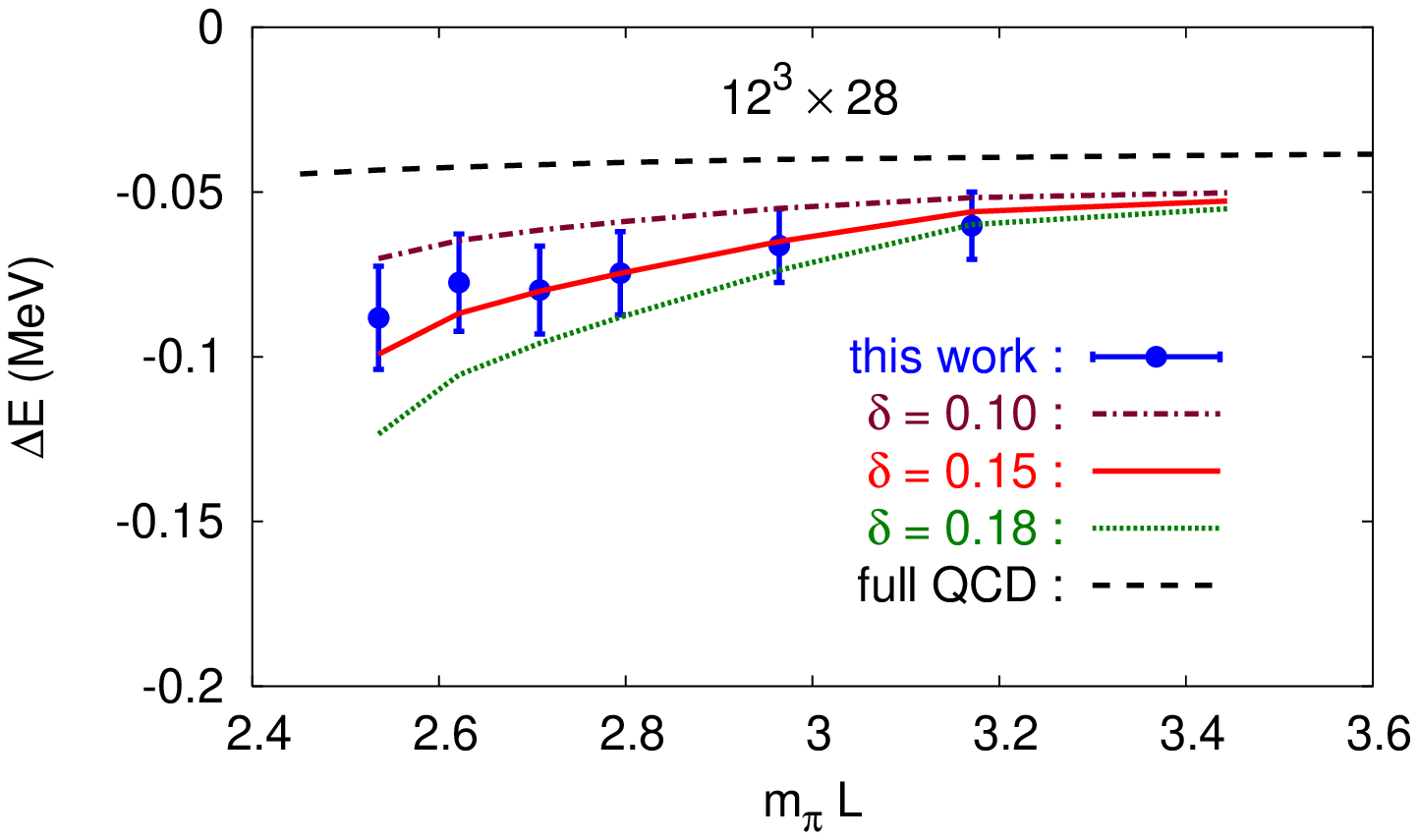}
  \includegraphics[width=0.5\hsize,height=0.15\vsize]{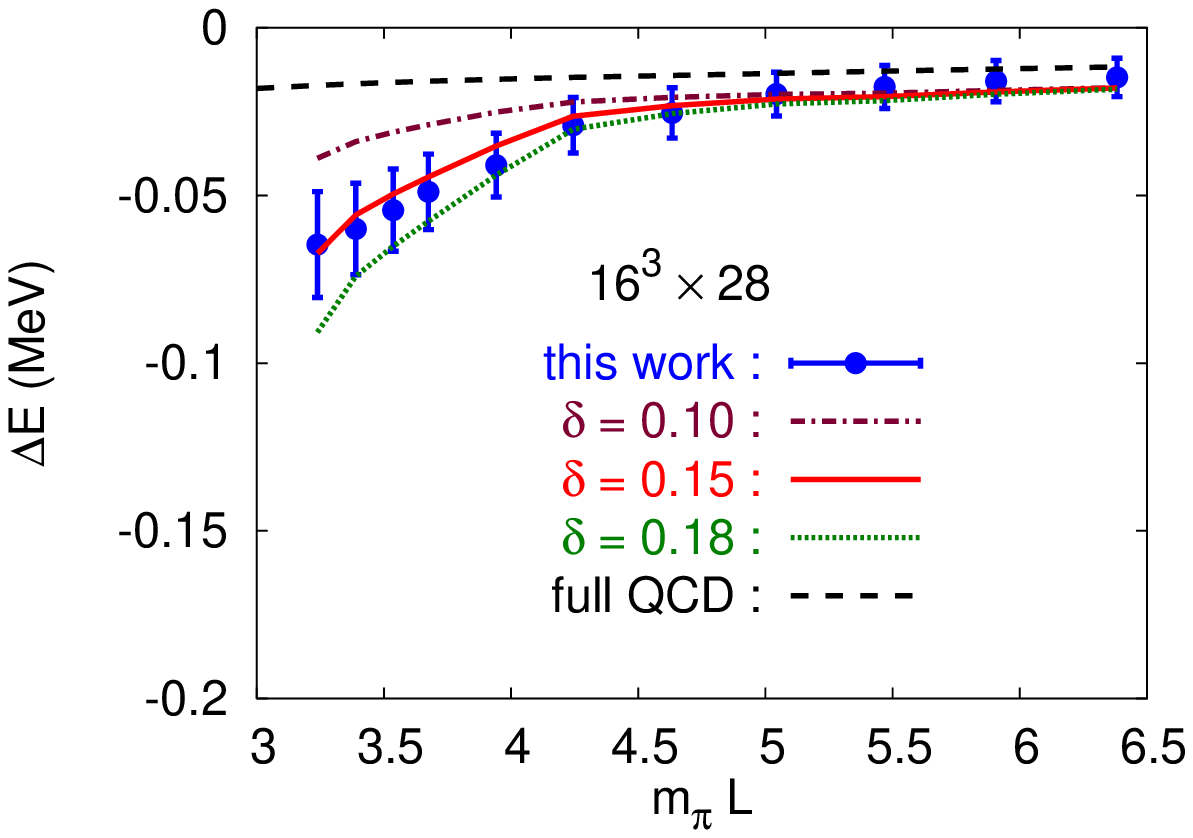}
  }
% \vspace*{-0.1in}
  \caption{The calculated energy shift of the lowest $I=0$ $\pi\pi$ state
           $\Delta E$ as a function of $m_{\pi}L$ for the $12^3 \times 28$
           lattice (left panel) and the $16^3 \times 28$ lattice (right panel).
           The lines labeled with $\delta$ indicate the quenched one-loop
           chiral perturbation results for $\delta= 0.10, 0.15$ and $0.18$.
           The one-loop full QCD chiral perturbation prediction is also given
           for comparison.  }
% \vspace*{-0.3in}
\label{deltaE}
\end{figure}

We interpolate $B_0(m_{\pi}L$) and $A_0(m_{\pi}L$) listed in Ref.~\cite{bg96}
for the range of $m_{\pi}L$ appropriate for our calculation on the $12^3 \times
28$ and $16^3 \times 28$ lattices and plot $\Delta E$ for $\delta = 0.10, 0.15$
and 0.18 which cover the range of $\delta$ corresponding to the
Witten-Veneziano formula for the $\eta'$ mass and from the study of quenched
chiral log in the pseudoscalar meson masses~\cite{ddh03a}.  The results are
presented in Fig.~\ref{deltaE} together with our data on the $12^3 \times 28$
lattice (left panel) and the $16^3 \times 28$ lattice (right panel) from the
lowest state in our calculation which we believe is the two-pion scattering
state.  We see that our data are reasonably consistent with the one-loop
quenched chiral perturbation calculation~\cite{bg96} for the range of $\delta$,
i.e.\ from 0.10 to 0.18, especially for $m_{\pi}L \ge 2.8$ for the $12^3 \times
28$ lattice and $m_{\pi}L \ge 3.5$ for the $16^3 \times 28$ lattice.  This is
true despite of the fact that we have not included the disconnected insertion
in our calculation.  This is consistent with our earlier finding that the
disconnected correlator is about an order of magnitude smaller than the
connected one at several time separations.

  We note that the tree-level energy shift in Eq.~(\ref{tree}) is the same in quenched 
and full QCD. If the energy shift from the loop is small compared to the
tree result in both the quenched and full QCD, then the quenched energy shift 
in Eq.~(\ref{energy-shift}) would agree with the full QCD case in Eq.~(\ref{full_QCD}). 
To this end, we plot the energy shift from Eq.~(\ref{full_QCD}) with the scattering length 
$a_0^{I=0} = \frac{7 m_{\pi}}{16\pi f_{\pi}^2}$ from the tree-level in Fig. \ref{deltaE}
and find that it is about a factor of two to three smaller than our data in the low $m_{\pi}L$
region. This shows that the loop contribution in the quenched case is enhanced compared
to that in full QCD and not small compared
to the tree part and thus one cannot apply the full QCD relation in Eq.~(\ref{full_QCD}) to
obtain the scattering length. One can only compare the energy shift to that of
the quenched chiral perturbation theory as was done in the last paragraph.

The third state in Fig.~\ref{sigma_pipi} with a large error bar is about 1 GeV
above the lowest state.  The fact that it is higher than the energy of the
non-interacting two-pion state (each pion with momentum $p_1=524$ MeV), as
indicated by the higher solid line, is an indication that the highest fitted
state is always higher than the true state as it inevitably includes the
unfitted higher states and hence cannot be taken as a good signal for a
definite state.

\begin{figure}[t]
% \vspace*{-0.1in}
  \centerline{%
% \hspace*{+0.04\hsize}
  \includegraphics[width=1.0\hsize]{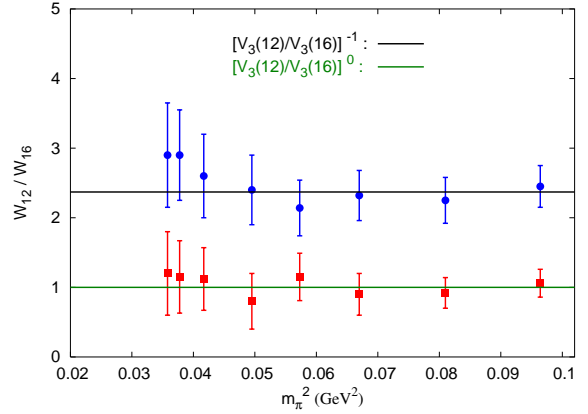}
  }
  \vspace*{-0.08in}
  \caption{Spectral weight ratio $W_{12}/W_{16}$ as a function of $m_{\pi}^2$
           for the lowest state (filled circle in Fig.~\ref{sigma_pipi}) and
           the next lowest state (filled square in Fig.~\ref{sigma_pipi}).   }
%  \vspace*{-0.2in}
\label{ratio}
\end{figure}

One interesting aspect of the spectrum is that there is an extra state between
the lowest $\pi\pi$ scattering state and the third state which presumably
encompasses the higher scattering states.  It has a sizable spectral weight, as
large as that of the lower $\pi\pi$ scattering state.  The mass is around 600
MeV.  It is tantalizing to identify it with $\sigma(600)$.  To verify this, we
study the volume dependence of the spectral weight of these states.  It was
advocated in the study of the Roper resonance, the pentaquark, and the ghost
state~\cite{mcd05,mla04} that one efficient way of distinguishing a
one-particle state from a two-particle scattering state in a finite box is to
study the volume dependence of its spectral weight.  From the normalization
factor of $1/\sqrt{V_3}$ for a particle in a box and the way the correlator is
constructed, i.e.\ with a point source and a zero-momentum sink, the spectral
weight of a one-particle state does not explicitly depend on volume; whereas,
the spectral weight of a weakly interacting two-particle state has an explicit
$1/V_3$ dependence~\cite{mla04}.  This is true when the one particle state is
reasonably separated from the scattering states so that the mixing of the two
states is not strong enough to spoil the characteristic volume dependence of
their respective spectral weights.  Since we have two lattices with sizes $12^3
\times 28$ and $16^3 \times 28$, the spectral weight ratio for a two-particle
state should be $W_{12}/W_{16} = V_3(16)/V_3(12) = 16^3/12^3 = 2.37$.  Plotted
in Fig.~\ref{ratio} are the ratios of the spectral weights for the lowest state
in Fig.~\ref{sigma_pipi} and the first excited state around 600 MeV.  We see
that the spectral weight ratio $W_{12}/W_{16}$ for the lowest state clusters
around 2.37, confirming our speculation that it is the interacting two pion
state.  On the other hand, the spectral weight ratio $W_{12}/W_{16}$ of the
excited state near $600$ MeV turns out to be consistent with unity.  This
suggests that this state is a one-particle state, not a $\pi\pi$ scattering
state, and strongly supports the identification of it to be the $\sigma(600)$.
Furthermore, by virtue of the fact that the ratios of spectral weights of these
two low-lying states are consistent with unity and $V_3(16)/V_3(12)$, it
suggests that the mixing of the two states, if any, is small.  In other words,
the two states with an energy separation of $\sim 300$ MeV are reasonably well
separated compared to half of the decay width of $\sigma(600)$.  Extrapolated
to the chiral limit, the mass of the one-particle state is $540\pm 170$ MeV.

    There have been concerns that the zero mode contributions which are finite 
volume effects may contaminate the results when the volume is small for the quark
mass under study. In the case of the chiral condensate, the zero modes can
be avoided by doubling the contribution from the chiral sector which does not
have zero modes~\cite{ehn99}. For the pion mass calculation, the zero modes can be
removed by considering the correlator of the pseudoscalar-pseudoscalar (PP) and 
scalar-scalar (SS) combination $\langle PP\rangle - \langle SS\rangle$
~\cite{bcc04,ddh02}. We have studied the zero mode contributions to the pion mass 
calculation on the same set of lattices used in the present study~\cite{ddh03a} 
by comparing the pion mass from the $\langle PP\rangle$ correlator and the
$\langle PP\rangle - \langle SS\rangle$ correlator and found that there is no 
difference in the pion masses within errors. We concluded that the volume is 
large enough for this lattice that the zero mode contribution is less than the 
error for the range of quark mass studied. In the present study of $a_0(1450), 
K_0^*(1430)$ and $\sigma(600)$, we don't see any indication of mass divergence 
for small quark masses. We thus believe that the zero mode contributions are within 
errors in the connected insertions. 

Although lattice study of tetraquark states started some time
ago~\cite{lll90,gmp93,aj00,chiu06}, we believe this is the first time that both
the one-particle state and its concomitant scattering state are identified and
their nature verified through the volume dependence of the spectral weights.
Short of such identification, we don't think one is able to rigorously confirm
the existence of the tetraquark mesonium.

Finally, we note that the calculation of tetraquark state with the $\bar{\psi}\gamma_5\psi
\bar{\psi}\gamma_5\psi$ interpolation field has first been attempted~\cite{aj00}.  It
was found that in the $I=0$ channel, the ground state is lower than the expected
two interacting pion state from chiral perturbation theory and is interpreted
as a bound state --- a tetraquark mesonium.  However, in this analysis the full
QCD formula in Eq.~(\ref{full_QCD}) was employed which we pointed out earlier
is not applicable to quenched calculations.  If the quenched chiral
perturbation calculation~\cite{bg96}, which has a different lattice length
dependence, is used one might come to a different conclusion.

\section{Conclusion}

To conclude, we calculated the isovector $a_0$ with the $\bar{\psi}\psi$
interpolation field and $\sigma(600)$ with the tetraquark interpolation field.
With the overlap fermion, we have come down in the chiral region with very low
pion mass ($182(8)$ MeV) in the quenched approximation.  After removing the
fitted $\eta' \pi$ ghost states, we found the lowest $a_0$ at $1.42 \pm 0.13$
GeV and $K_0^{*}$ at $1.41 \pm 0.12$ GeV which are consistent with the
experimental $a_0(1450)$ and $K_0^{*}(1430)$ being the $q\bar{q}$ states and
confirms the earlier findings in quenched and partially quenched calculations
at higher quark masses.  In addition, we have been able to fit the $I =0$
scalar tetraquark correlator and have identified, through the volume study of
their spectral weights, both the lowest interacting two-pion state and a
one-particle state at $540 \pm 170$ MeV.  This suggests that $\sigma(600)$ does
exist as a particle and it is a tetraquark mesonium.  This is consistent with
the recent dispersion relation analysis of $\pi\pi$ and $K\overline{K}$
scattering with the Roy equation which has led to the $\sigma$ pole at
$441_{-8}^{+16} $ MeV with a width $\Gamma_{\sigma} = 544_{-25}^{+18}$
MeV~\cite{ccl06}.  Further lattice calculations with dynamical fermion in the
chiral region with $m_{\pi} < 300$ MeV are needed to check these results.

This work is supported by JSA, LLC under U.S. DOE Contract No.
DE-AC05-06OR23177 and by DOE grant DE-FG05-84ER40154, and NFSC grants 10575107
and 10675101. We wish to thank G.E. Brown, M. Chanowitz, H.Y. Cheng, M. Golterman, 
R. Jaffe, C. Liu, M. L\"{u}scher, D.O. Riska, and F. Wilczek for useful discussions.

%\vspace*{-0.76cm}

\end{document}